\documentclass[twocolumn,prb,preprintnumbers,amsmath,amssymb]{revtex4}
\usepackage[pdftex]{graphicx}
\usepackage{epsfig}
\usepackage{psfrag}
\usepackage{upgreek}
\usepackage{wrapfig}

\newcommand{\ahl}{$\alpha$HL }
\begin{document}

\title{Rapid Sequencing of Individual DNA Molecules in Graphene Nanogaps}

\author{Henk W.Ch.~Postma}
 \homepage{http://www.csun.edu/~hpostma/}
\email{postma@csun.edu}
\affiliation{
Department of Physics, California State University Northridge, 
18111 Nordhoff Street, Northridge, CA 91330-8268
}

\date{\today}

\begin{abstract}
I propose a technique for reading the base sequence of a single DNA molecule using a graphene nanogap. 
\end{abstract}
 

\maketitle


One of the greatest challenges of biotechnology is establishing the base sequence of individual molecules of DNA without the need for PCR amplification or other modification of the molecule. The Sanger method \cite{sanger_dna_1977} has proven extremely powerful and has resulted in the recent sequencing of the human genome in a monumental collaborative effort. \cite{lander_e.s._initial_2001,venter_sequence_2001}

Sequencing human DNA occurs through {\em shotgun sequencing}\cite{venter_sequence_2001,lander_e.s._initial_2001} which is a strategy around the technique introduced more than 30 years ago by Sanger {\em et al.} \cite{sanger_dna_1977} It consists of breaking the sample into small random fragments and amplifying them, sequencing these fragments using the Sanger method, and merging these sequences by determining overlapping areas by their base sequence. There are many challenges to making current sequencing technology more cost effective and comprehensive. 1) The total process is time and resource intensive because the Sanger read length is short, requiring many small sequencing steps, many overlapping reads, and a lot of computational power to merge the sequences. 2) DNA amplification is required. Bacterial cloning with E. Coli sometimes contaminates read sequences with bacterial material. PCR sometimes creates artificially long repetitive segments due to polymerase stuttering \cite{bilsel_polymerase_1990}, or merges two unrelated sequences thereby creating a DNA segment that does not occur in the original sequence. In addition, it is a time and cost-intensive process and since it is at the heart of the sequencing process, it quickly increases the overall cost and time required for whole-genome sequencing. 3) The samples need to be tagged with fluorescent or radioactive labels to image the DNA fragments after gel electrophoresis. 4) It is not possible to sequence large homopolymeric segments, e.g. telomeres, of the genome due to the finite Sanger read length. 

Using the requirement of the X-prize, to sequence 100 genomes in 10 days \cite{archon}, as a benchmark for future sequencing technology with a single device that will sequence all of these genomes sequentially, without any pre- or post-processing, a $\sim 3 \mu$s read time per base is required.

Numerous improvements are being developed, optimizing various aspects of the sequencing process. \cite{shendure_advanced_2004,fredlake_what_2006}  Miniaturization with microfluidics is being developed to improve the readout speed, reduce the volume of material needed, and reduce the cost per base sequenced,  while still relying on the proven Sanger method. \cite{emrich_microfabricated_1993,koutny_eight_2000} Also, reversible terminators are being developed which will allow for sequencing of homopolymeric sequences. \cite{metzker_termination_1994,welch_synthesis_1999} Finally, several single-molecule sequencing techniques are being developed. These represent a different strategy that deviate from the Sanger method. They require very little genome material and therefore no amplification. One such method demonstrated single-nucleotide microscopy of fluorescently labeled nucleotides that were inserted into individual DNA molecules. \cite{braslavsky_sequence_2003}

{\em Nanopore-based sequencing} is a single-molecule sequencing technique that is especially promising. It is believed that a large read length and high throughput can be achieved simultaneously.\cite{zwolak_colloquium:_2008} The first translocation studies of individual DNA molecules were conducted with naturally occurring alpha-hemolysin (\ahl) proteins that spontaneously embed themselves in a lipid bilayer and form a nanopore. This \ahl pore is studied using electrophysiology, in which a patch-clamp amplifier records the current through the protein pore while a DNA molecule translocates through it under the influence of an applied transmembrane electric field acting on the negatively-charged backbone. \cite{henry_blockade_1989,bayley_triggers_1994,bezrukov_counting_1994,bustamante_patch_1995,kasianowicz_characterization_1996,bezrukov_dynamics_1996,howorka_sequence-specific_2001,heng_electromechanics_2006,zhao_single-strand_2007} Both single-stranded DNA (ssDNA) and double-stranded DNA (dsDNA) have been studied. The minimum pore size that ssDNA can translocate through is 1.5 nm\cite{zhao_single-strand_2007} while it is 3 nm for dsDNA. \cite{heng_electromechanics_2006} 

Biological nanopores and the lipid bilayer membrane they are embedded in are only stable within a small range of temperature, pH, chemical environments, and applied electric fields, limiting practical applications. {\em Solid-state} nanopores do not suffer from this. Solid-state nanopores have been fabricated in Si$_3$N$_4$ membranes \cite{li_ion-beam_2001}, SiO$_2$ membranes \cite{storm_fabrication_2003}, and polymer films\cite{mara_asymmetric_2004}. Translocation studies of dsDNA showed very high velocities\cite{storm_fast_2005}, owing to the much reduced interaction of DNA with solid-state nanopores as compared to \ahl pores \cite{zhang_effective_2007,hu_theory_2008}. 

{\em Nanopore-based sequencing} using a transverse conductance measurement of a DNA molecule while it translocates through the nanopore has been suggested as an alternative to the Sanger method. \cite{zwolak_electronic_2005,lagerqvist_fast_2006} The idea is that different bases have different local electronic densities of states with different spatial extent owing to their different chemical composition. If the bases are passing through a voltage-biased tunnel gap one by one, they will periodically alter the current based on whether the localized states in the bases are contributing to the tunnel current. Analyzing the current as a function of base is then expected to reveal the base sequence. However, making nano-electrodes that are aligned with the nanopore is very challenging. 


\begin{figure}
\begin{center}
\includegraphics [width=8.0 cm, height= 7.0 cm]{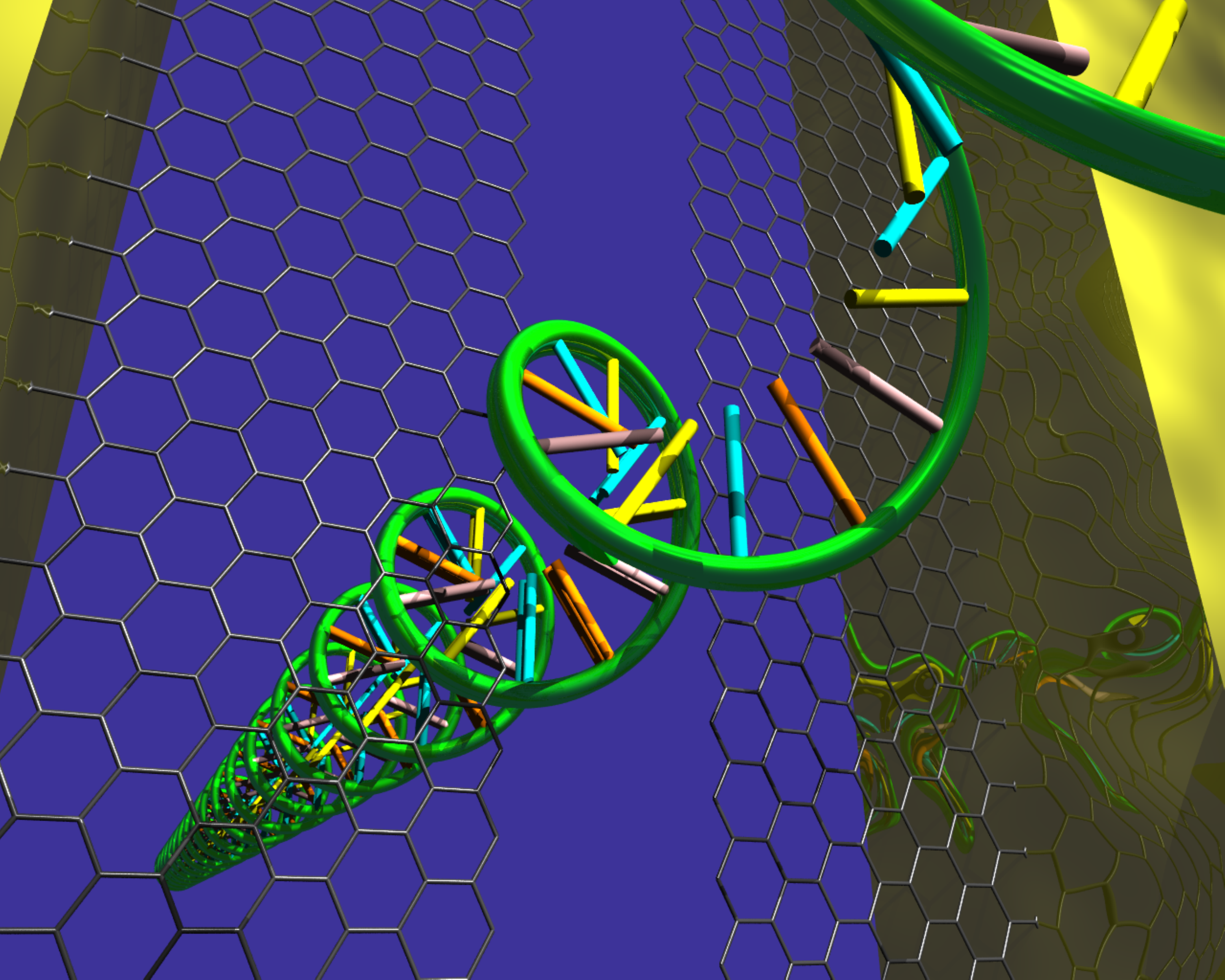}
\caption{ \label{fig1} {\sl The individual bases of a ssDNA molecule (backbone in green, bases in alternating colors) sequentially occupy a gap in graphene (hexagonal lattice) while translocating through it. Their conductance is read, revealing the sequence of the molecule. The contacting electrodes to the graphene nanogap (Au, yellow) are on the far left and right side of this image. } 
}
\end{center}
\end{figure}

Here, I propose to use graphene nanogaps for DNA sequencing, using the graphene as the electrode as well as the membrane material. The experimental layout is drawn in figure \ref{fig1}. Graphene, a single-atom thick hexagonal carbon lattice that has recently been discovered \cite{novoselov_electric_2004}, can be synthesized in a variety of manners. \cite{berger_ultrathin_2004,gilje_chemical_2007,li_chemically_2008} It is an ideal material for making nanogaps for sequencing due to its single-atom thickness $d$, its ability to survive large transmembrane pressures \cite{bunch_impermeable_2008,poot_nanomechanical_2008,lee_measurement_2008}, and its intrinsic conducting properties. The last property is especially advantageous because the membrane {\em is} the electrode, automatically solving the problem of having to fabricate nanoelectrodes that are carefully aligned with a nanogap. Contacts to the graphene sheet can be fabricated using standard electron-beam lithography, metal evaporation and lift off. The graphene sheet is covering a $\sim 500$ nm micropore in a Si/SiO2 wafer and the wafer is mounted in a PDMS fluid cell with integrated Ag/AgCl electrodes for ion current measurement as well as contacts to the Au electrodes for transverse conductance measurement.
 
Various methods can be used to obtain graphene nanogaps.  They may be fabricated by nanolithography with a scanning tunneling microscope (STM), in a method similar to that used for cutting carbon nanotubes. \cite{venema_length_1997,rubio_a._mechanism_2000} Recently, STM nanolithography on the top graphene layer of grapite was demonstrated. \cite{tapaszto_tailoring_2008} The ideal nanogap width is $1.5 - 2.0$ nm, allowing for ssDNA to pass through it in an unfolded state \cite{zhao_single-strand_2007}  as well as assuring the largest transverse current. The transverse conductance of DNA molecules can then be measured while they translocate through a nanogap in the graphene membrane, revealing the base sequence of the molecule.

The DNA translocation speed is typically much larger in solid-state nanopores
than in biological nanopores, owing to their large difference in size and
aspect ratio. 
\cite{storm_fast_2005,chauwin_strong_1998,ghosal_electrokinetic-flow-induced_2007}
For pore sizes that are small compared to the ssDNA width, the bases stick to
the side of the nanogap, lagging behind the backbone, while the molecule moves
through the gap. \cite{sigalov_detection_2008} For large gap sizes, the bases'
orientation can vary significantly, but they can be aligned by the electric
field due to the applied bias voltage $V_{bias}$ across the
gap. \cite{lagerqvist_fast_2006,lagerqvist_comment_2007}

When large ($\sim 10 - 100$ kbp) dsDNA translocates through solid-state
nanopores with a diameter much wider than the molecule, the velocity depends as
a power-law on the length 
\[ v \propto L^{2\nu-1} = L^{-0.27} \] 
where $\nu=0.611$ is the Flory exponent\cite{smith_dynamical_1996} and the required applied
electric field strength is relatively low, $E = 6.0 \times 10^{6}$ V/m.
\cite{storm_fast_2005} In contrast, 'long' ($\gg 12$
nt) ssDNA translocates through a much more narrow (1.8 nm) and $d=5.2$ 
nm deep \ahl nanopore with length-indepent velocity. The velocity
depends quadratically on a much larger required driving voltage $V$ as 
\[ v = k_1 \left(V-\frac{E_0}{d}\right)^2 + k_2
\qquad , 
\] where $E_0/d = 47$ mV, $k_2 = 0.006$ nm/us, and $k_1 = 2.0$
nm/usV$^2$.
\cite{meller_voltage-driven_2001} The electric field threshold for DNA translocation $E_0$, depends on the pH and pore geometry and is due to a stretching transition of the molecule into the pore. \cite{heng_electromechanics_2006,zhang_effective_2007}

The \ahl pore geometry is very close to that proposed here, since 1) the
ideal graphene nanogap width of 1.5 nm is similar and 2) the
narrowest region of the \ahl pore and the graphene nanogap are
similar in thickness.  This may result in similar DNA-graphene nanogap interaction
strengths although a full model is required. \cite{gowtham_physisorption_2007} An advantage of graphene nanogaps is then that their local atomic configuration
can be imaged directly with the STM after the gap has been fabricated allowing
for a comprehensive comparison of measurements with theoretical calculations.
Assuming an average field strength in the \ahl pore of $250
\mbox{ mV}/5.2 \mbox{ nm}=48 \mbox{MV}/\mbox{m}$, we can extrapolate that an applied voltage of 30 mV across
the graphene membrane with effective thickness of 0.6 nm will yield an average
translocation time of 3.6 us/nt. The voltage that is applied across the nanogap to read the DNA's transverse conductance is expected to slightly alter the translocation velocity. \cite{hu_theory_2008}

It has been suggested that the conduction mechanism that allows one to
distinguish between the different bases depends on the spatial extend of the
HOMO and LUMO levels (which are typically far away from the fermi level of the
leads) and their overlap with the electrode
wavefunction. \cite{zwolak_electronic_2005,lagerqvist_fast_2006,zwolak_colloquium:_2008}
More recently, it was found that poly(GC) and poly(AT) can be distinguished
electronically through measurement of localized states around $V_{bias} = 0$. \cite{xu_electronic_2007} One can then estimate the current due to the bases by evaluating 
\[
	I = A \int D_L(E) D_R(E-eV_{bias}) |T(E)|^2 \mathrm{d}E
\] where $T(E)$ is the effective transmission of the electronic base states, and $D_{L,R}$ are the densities of states of the left and right electrodes, respectively. \cite{shapir_electronic_2008} For a realistic
description of the tunnel current in the proposed experiment, both the distance dependence for this resonant-tunneling regime\cite{payne_transfer_1986}, counter
ions \cite{shapir_electronic_2008}, and the unique density of states of graphene\cite{novoselov_electric_2004,berger_ultrathin_2004,gilje_chemical_2007,li_chemically_2008} need to be taken into account. Future studies of this system will also need to include the doping due to adsorbed water molecules on the graphene membrane and its reduction in the absense of an underlying SiO$_2$ substrate. \cite{wehling_first-principles_2008}

Although preliminary experiments indicate otherwise\cite{xu_electronic_2007},
it has been argued that transverse electronic transport cannot be used for DNA sequencing. \cite{zhang_first-principles_2006} As an
alternative, the  layout proposed here can also be used to directly detect
voltage fluctuations due to the local and unique dipole moments of the
bases. \cite{gracheva_electrical_2006,gracheva_simulation_2006} This capacitive detection approach is not
preferred, however, due to its reliance on the relatively long-range
capacitive interaction, possibly limiting the spatial resolution with which
individual bases can be resolved.

\noindent {\bf Acknowledgments:} 
I thank Michael Dickson, Karapet Karapetyan, Hanyu Lee, George Gomes, Konstantin Daskalov, Michael Zwolak, and Marc Bockrath for discussions.

\end{document}